\documentclass[]{spie}  %>>> use for US letter paper
\usepackage{amsmath,amsfonts,amssymb}
\usepackage{algorithm}
\usepackage[noend]{algpseudocode}
\usepackage{graphicx}
\usepackage{textcomp}\usepackage[colorlinks=true, allcolors=blue]{hyperref}

\title{Point Source Localization with a Planar Optical Phased Array Compressive Sensor}

\author[a,b]{Julian A. Brown}
\author[a]{Steven J. Spector}
\author[a]{Michael Moebius}
\author[a]{Lucas Benney}
\author[a]{Daniel Vresilovic}
\author[a]{Brian Dolle}
\author[a]{Alexandra Z. Greenbaum}
\author[a,d]{Alex Huang}
\author[e]{Christopher V. Poulton}
\author[e]{Michael~R.~Watts}
\author[a]{Robin Dawson}
\author[a,c]{Benjamin F. Lane}
\author[a]{J.P. Laine}
\author[b]{Kerri Cahoy}
\author[a]{Hannah~A.~Clevenson}

\affil[a]{The Charles Stark Draper Laboratory, Cambridge, MA 02139, USA}
\affil[b]{MIT, Cambridge, MA 02139, USA}
\affil[c]{Current Affiliation: Systems \& Technology Research, Woburn, MA 01801, USA}
\affil[d]{University of Massachusetts, Amherst, MA 01003, USA}
\affil[e]{Analog Photonics, Boston, MA 02210, USA}

\authorinfo{Further author information: (Send correspondence to H.A.C.)\\H.A.C.: E-mail: hclevenson@draper.com}

\begin{document} 
\maketitle

\begin{abstract}
Compressive sensing has been used to demonstrate scene reconstruction and source localization in a wide variety of devices.  To date, optical compressive sensors have not been able to achieve significant volume reduction relative to conventional optics of equivalent angular resolution.  Here, we adapt silicon-photonic optical phased array technology to demonstrate, to our knowledge, the first application of compressive imaging in a photonic-integrated device.  Our novel sensor consists of an $8\times 8$ grid of grating couplers with a spacing of $100~\mu$m.  Path-matched waveguides route to a single multimode interferometer (MMI), which mixes and randomizes the signals into 64 outputs to be used for compressed sensing.  Our device is fully passive, having no need for phase shifters, as measurement matrix calibration makes the measurements robust to phase errors.  For testing, we use an Amplified Spontaneous Emission (ASE) source with a bandwidth of 40 nm, centered at 1545 nm.  We demonstrate simultaneous multi-point (2 sources demonstrated in this work)  brightness recovery and localization with better than 10 arcsecond precision in a sub-millimeter thick form-factor.  We achieve a single source recovery rate higher than 99.9\% using 10 of the 64 outputs, and a 90\% recovery rate with only 6 outputs, 10 times fewer than the 64 needed for conventional imaging.  This planar optical phased array compressive sensor is well-suited for imaging sparse scenes in applications constrained by form factor, volume, or high-cost detectors, with the potential to revolutionize endoscopy, beam locators, and LIDAR.  
\end{abstract}


\begin{figure}[ht]
	\begin{center}
		\includegraphics[height=5cm]{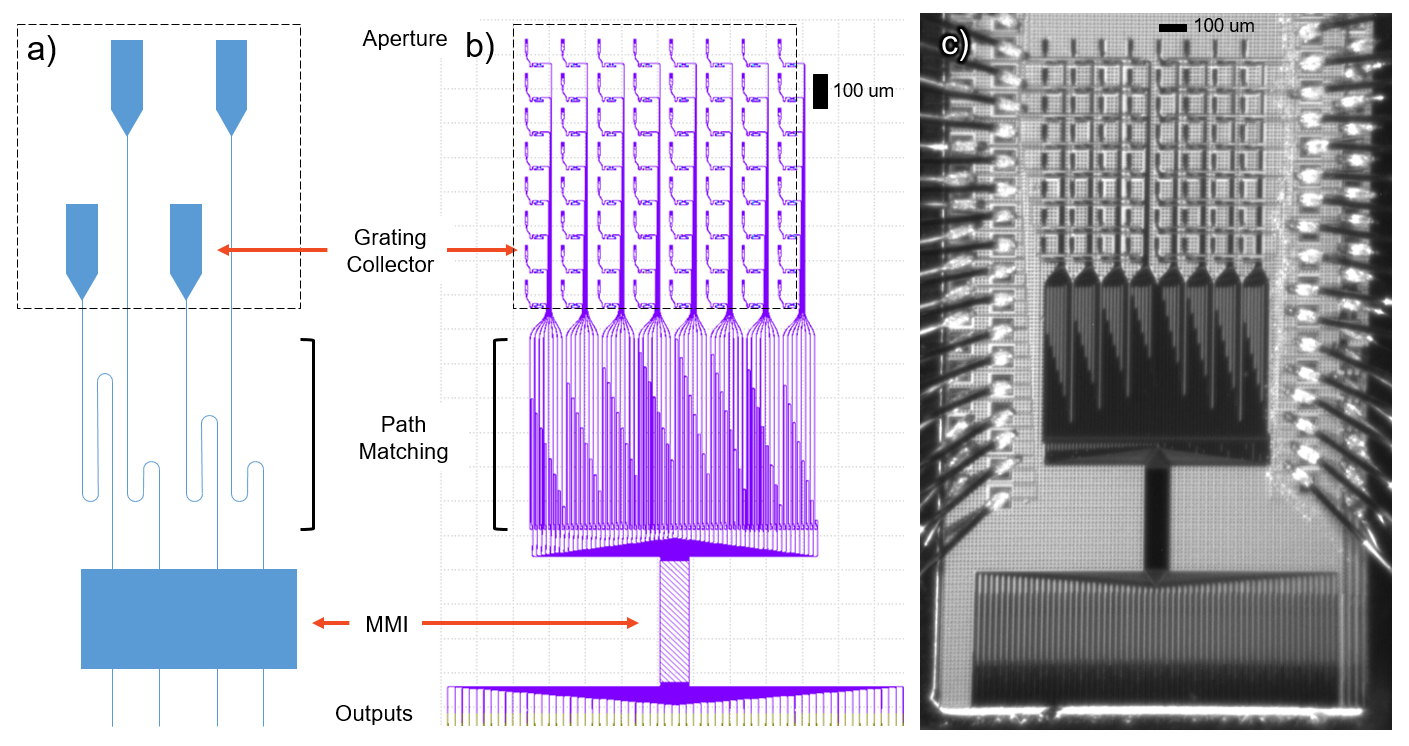}
	\end{center}
	\caption{\textbf{Device Layout} – a) A simplified diagram of our device layout.  A grid of grating couplers act as antennas, collecting light perpendicular to the grid and redirecting it into waveguides.  To minimize wavelength-dependence, the waveguides connecting the grating couplers to the MMI are path-matched with variable-length S-bends.  The MMI’s waveguide outputs are edge-coupled for external detection.  b) Our device’s actual mask design used on the wafer.  c) A close-up of a wafer containing our device.}
	\label{fig:device_layout}
\end{figure}

\section{INTRODUCTION}
\label{sec:intro}

Compressive sensing is a sensing modality that allows signals to be severely under-sampled and later reconstructed\cite{8260873}. This allows for simpler hardware, lower power consumption, and lower memory requirements, making it well-suited for remote and distributed sensing tasks. To date, however, optical compressive sensors have not achieved significant volume reduction relative to conventional optics of similar angular resolution.  

Methods for miniaturizing imaging systems are sought in fields ranging from astronomy to portable consumer devices to endoscopy\cite{britten_large-aperture_2014, byrnes_designing_2016, tragardh_label-free_2019, shin_single-pixel_2016}. However, the system aperture size fundamentally limits imager performance, determining both the light collection area and angular resolution. Conventional imaging systems rely on lenses or mirrors to focus light and form an image on a detector array. These imaging systems have focal lengths comparable to the aperture diameter, so the system volume scales with the aperture diameter cubed \cite{ray_applied_2002}. A folded optical system can reduce the depth of the imager \cite{li_ultrathin_2016} at the expense of light collecting area.

Micro-lithographic techniques, developed in the microelectronics industry, have enabled the fabrication of complex planar optical systems, including imaging interferometers and phased-array imaging systems typically implemented at radio frequencies.  Silicon photonics research is largely focused on applications in the telecom and datacom industries \cite{inniss_silicon_2016,nagarajan_inp_2010, alduino_demonstration_2010,chen_silicon_2016,zhang_8_2016}. Work has also been done in beamforming using optical phased arrays (OPAs) \cite{kwong_-chip_2014,abediasl_monolithic_2015, zadka_millimeter_2017, clevenson_incoherent_2020}.

Prior work has shown that multimode interferometers (MMIs) in integrated photonic devices can create a sufficiently random measurement matrix for use in a compressive sensor\cite{valley_multimode_2016}.  MMIs are broad waveguides, which act as optical mixing regions.  Light is guided in multiple single-mode waveguides, which are brought together to allow the light to interact within the MMI before being split back up into single-mode waveguides.  In theory, this mixing can be "lossless"\cite{heaton_general_1999}.  Heaton and Jenkins considered compact, approximately square MMIs, with lengths and widths both proportional to their number of ports.  They showed that these MMIs are capable of distributing light from each input approximately evenly across their outputs.  Additionally, each output has a different phase relationship with each input, scrambling the input signals together.  Altogether, Heaton and Jenkins' MMI design is low-loss, compact, and random, making it an ideal choice for the mixing component in our compressive sensing device.

In this work, we design and fabricate a device based on a broadband, planar optical phased array architecture with an MMI element to mix and randomize the signal which enables its use as a compressive sensor. We then demonstrate point source localization using this silicon-photonic compressive sensor for both one and two sources in the field of view. We recover the source brightness and localize two light sources in the field of view with better than 10 arcsecond precision in a sub-millimeter thick form-factor. We achieve a single source recovery rate higher than 99.9\% using 10 of the 64 outputs, and a 90\% recovery rate with only 6 outputs, approximately 10 times fewer than the 64 needed for conventional imaging.  In an ideal scenario, the number of samples needed in a compressive sensor scales as the logarithm of the number of equivalent conventional outputs.  Data reduction of more than 100x are possible for mega-antenna class sensors.

\begin{figure}[ht]
  \begin{center}
  \includegraphics[height=6cm]{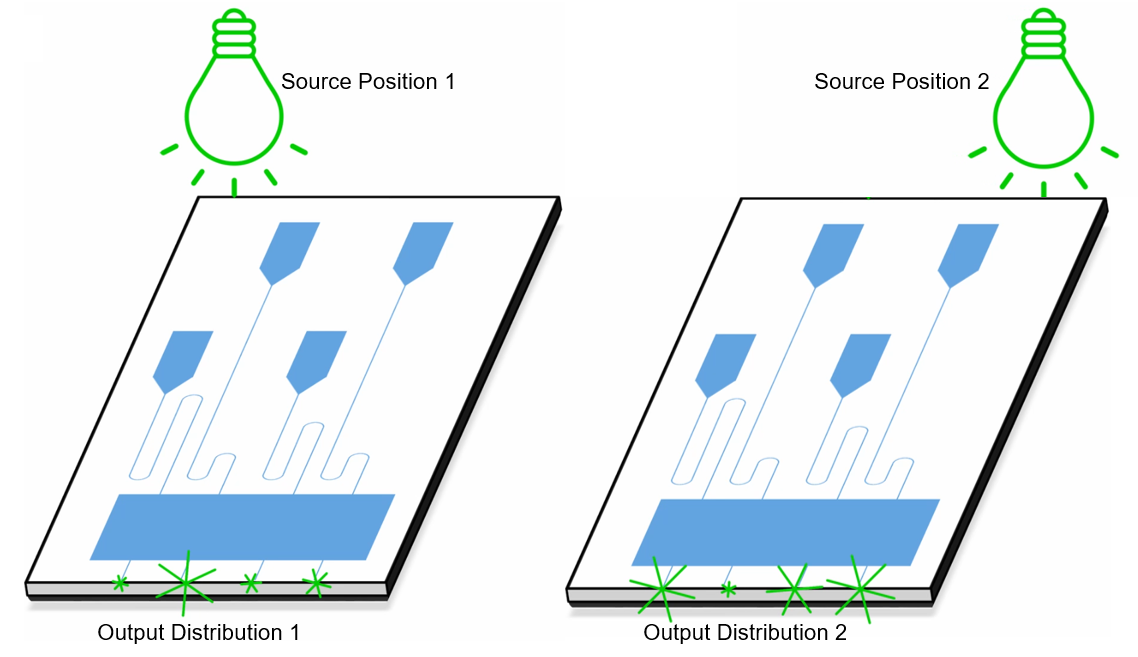}
  \end{center}
  \caption{\textbf{Theory of Operation} – The distribution of intensities at our device’s outputs depends on the angle of incident light.  We can create a lookup table mapping output intensity distributions to their angles of incidence.  Then, given some distribution of output intensities, we can determine which direction the light must have come from.}
  \label{fig:theory_of_operation}
\end{figure}

\section{Approach}

Our device consists of several key components: grating couplers, pathlength-matched waveguides, a mixing region (MMI), and outputs. Figure~\ref{fig:device_layout}(a) shows a simplified diagram of our sensor, (b) the mask design), and (c) a micrograph of the actual device. Light from the sources is coupled into integrated silicon waveguides by an $8\times 8$ square grid of grating couplers on a $100~ \mu$m pitch. Path-matched waveguides route the light from the grating couplers to the MMI, where the light is mixed and randomized, and then to the output waveguides, enabling broadband compressive-sensing imaging.

As the source moves across the field of view, the angle of the incident light changes. Figure~\ref{fig:theory_of_operation} illustrates the device concept: a unique and repeatable output distribution is formed for every source position in the field of view. These output distributions are calibrated to form a lookup table, allowing the recovery of the source position. Two sources in the field of view present a linear combination of lookup table elements. 

%Each waveguide includes an integrated resistive heater, which can adjust the phase delay of light in that waveguide. These phase shifters were not used for the result described here. 

The number of grating collectors along the y-axis of the imager is limited by the space required for path-matching. Gratings have a small form factor ($7\times10~\mu$m$^{2}$) to ensure broadband response. As a result, the fill factor of grating collectors is less than $1\%$, greatly limiting the effective aperture over which light is collected by the device. A microlens array  with $100~\mu$m pitch along the x- and y-axes can be aligned above the grating collectors, with the gratings at the lenslet focal plane, to improve the effective fill factor of the device by $\sim 80\times$. The lenslet array suppresses higher order grating lobes from the array of collectors, limiting light collection to the central grating lobe and yielding a 0.9° field of view at 1545 nm wavelength. As optical power and grating lobes were not potential issues during testing and to reduce confounding variables, the lenslet array was not used in our experiments.
 
Photonic component designs were initially generated using eigenmode expansion and then verified and optimized in finite difference time domain (FDTD) software. Components include grating couplers, waveguides, waveguide tapers, MMIs, and inverse-taper edge couplers. MMI designs were generated using analytical solutions \cite{Bachmann:94} and optimized using eigenmode expansion and variational 2.5D FDTD electro-magnetic field solvers (Ansys/Lumerical EME and varFDTD). Photonic chip layout was completed using Cadence Virtuoso. Devices were fabricated in a commercial complementary metal-oxide-semiconductor (CMOS) foundry on 300 mm silicon-on-insulator wafers with $2-3~\mu$m thick buried oxide. The fabrication process included the passive silicon waveguide layer, ion implants for resistive heaters (unused in this work), metal routing and contact layers for active control of heater elements, and cladding layers to ensure optical isolation between layers. 

\begin{figure}[ht]
  \begin{center}
  \includegraphics[height=6cm]{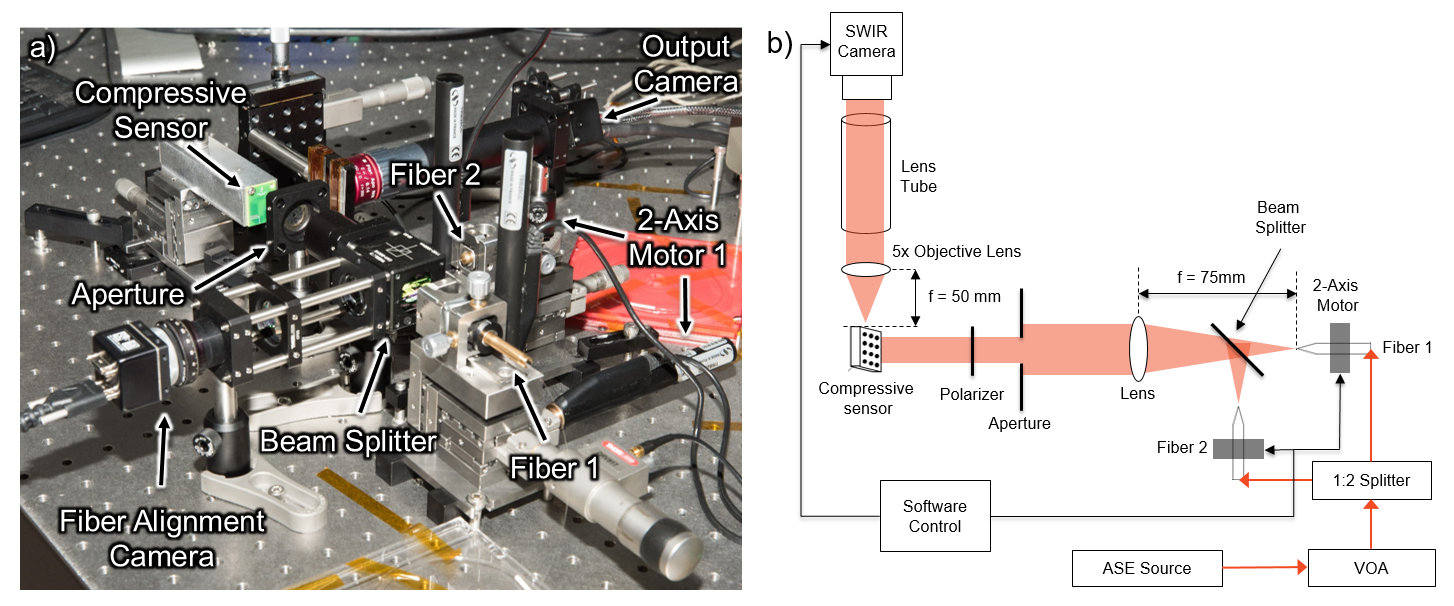}
  \end{center}
  \caption{\textbf{Experimental Setup} – a) Our optics bench two source test setup.  b) A block diagram of the same setup.  Point sources at infinity are simulated with fibers emitting light from an ASE source.  The fibers are mounted on stages controlled with TRB25CC actuators, allowing software control of the point sources’ x and y positions.  One fiber was disabled for single source testing and measurement matrix calibration.  Images of our device’s edge-coupled outputs are then used to reconstruct the point sources’ positions in software.}
  \label{fig:experimental_setup}
\end{figure}

In this work, we consider two cases: one or two point sources in the field of view. The experimental setup is the same for both, with only one source used in the former (Fig~\ref{fig:experimental_setup}).  Two point sources are constructed using broadband light from a C-band amplified spontaneous emission (ASE) source coupled into a variable optical attenuator (VOA). The signal is split equally into two SMF-28 optical fibers, each mounted on 3-axis fiber alignment stages. The stages consist of a standard micrometer for the z-axis and stepper motors (Newport TRB25CC) to control the fibers’ x- and y- positions. The light from the fibers is co-boresighted,  directed through a beam splitter, and collimated with a bi-convex spherical lens (d = 25.4 mm, f = 75 mm). The collimated beams travel through an aperture and a polarizer to decrease background light and to select the s-polarization, for which the grating couplers on the device are designed.  The readout of the device is done by re-imaging the light from the edge couplers with a 5x objective (Mitutoyo model M Plan Apo NIR 5X) onto a short wave infrared (SWIR) camera (Sensors Unlimited model SU640CSX). 

\begin{figure}[ht]
	\begin{center}
		\includegraphics[height=7cm]{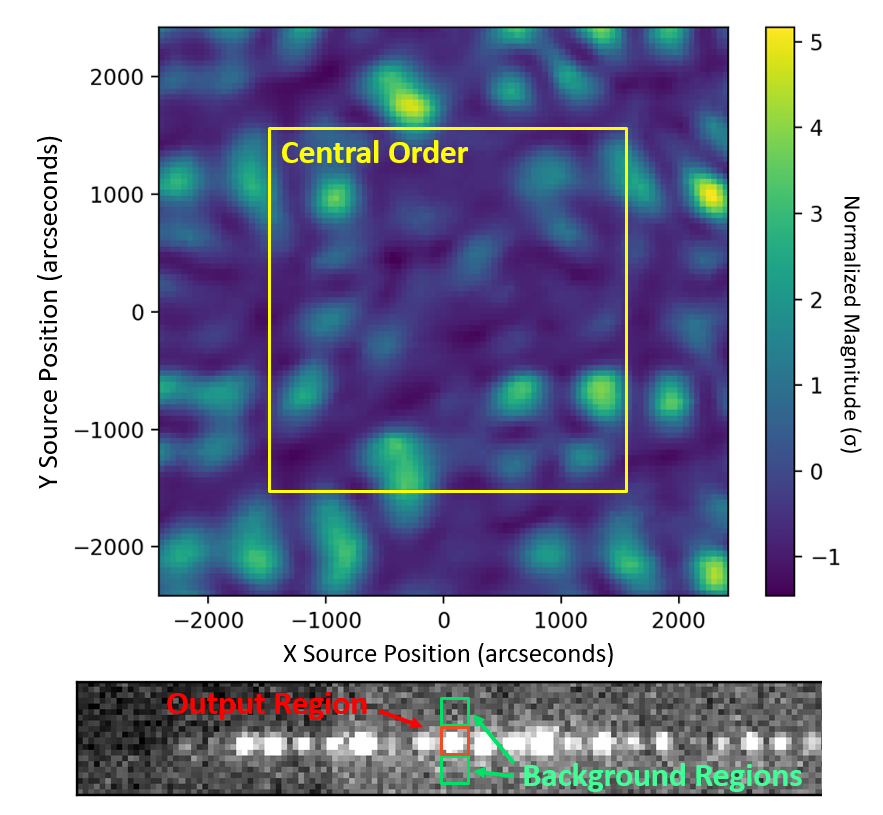}
	\end{center}
	\caption{\textbf{Speckle Calibration} – a) Variations in the brightness of the tenth output are plotted against the position of a single source.  The random fluctuations in intensity form a speckle pattern, which repeats outside of the central order.  b) The regions used for extracting output brightnesses are overlaid on a raw image of the edge couplers.  Each output’s brightness is the sum of a 5x5 pixel area around the output minus its background brightness.  To account for variations in background noise resulting from stray light, the background brightnesses are calculated as the average of the sums of the vertically adjacent 5x5 pixel areas.}
	\label{fig:speckle_calibration}
\end{figure}

The device edge couplers are imaged as the one or two point sources in the device’s far-field are varied in position (x and y), changing the angles of the collimated beams reaching the device.  As the source positions are scanned, the intensity pattern of the output couplers changes.  For each position, the data must be converted from an image into a vector of intensities.  The pixel locations of the edge couplers in the image are determined using a matched filter, and the intensity of each spot is calculated by the sum of the counts in a 5x5 pixel box (Fig.~\ref{fig:speckle_calibration}b).  To remove the effect of stray light, a simple background subtraction is performed, with the background for each spot calculated by the average of the sums of 5x5 pixel boxes above and below.  For a given output, the intensity varies randomly with source position within the central order of the device (Fig.~\ref{fig:speckle_calibration}a).  The pattern repeats in the higher orders (grating lobes outside the device FOV), so the analysis focuses only on the central order.

\section{Analysis}

Once the device has been calibrated, we need to select a compressive sensing algorithm before we can start recovering point sources.  In this paper, we compare two compressive sensing algorithms: the brute-force L0-norm and its linear relaxation L1-norm.  While the L0-norm boasts superior performance, the L1-norm is significantly faster for larger problems.

\subsection{L0 - Exhaustive Search}

The L0-norm itself is not an algorithm but an optimization constraint.  In our case, we find the least-squared-error recovery subject to an L0-norm constraint on $x$:
\begin{equation}
\label{eq:l0}
\begin{aligned}
\min_{\hat{x}} \|y-A\hat{x}\|_{2}^{2}
\quad 
\textrm{s.t.}
\quad 
\|\hat{x}\|_{0} \leq k
\end{aligned}
\end{equation}
where $y$ is the vector of background-subtracted test measurements, $A$'s columns are the background-subtracted calibration measurements, $k$ is the number of non-zero elements in $x$, and $\hat{x}$ is the recovered vector with length $n$.  $\hat{x}$'s non-zero elements give both the approximate positions of the sources and their brightnesses relative to the calibration source.  We use the naive algorithm, which recovers $\hat{x}$ by iterating over every combination of non-zero elements, solving the now well-conditioned inverse sub-problems with the Moore-Penrose inverse.  $\hat{x}$ is then the sub-problem solution with the smallest error\cite{foucart_mathematical_2013}.  Our solution is guaranteed to be optimal for calibration grid-aligned test sources under Eq.\eqref{eq:l0}, as it checks every possible assignment of $\hat{x}$.\footnote{We did not implement the continuous-space extension, but it only requires a small addition.  Test sources between grid points can be recovered by running gradient descent over an interpolation of the calibration grid, such as Levenberg-Marquardt over a cubic-spline.  When ambiguous global minima are possible, the gradient descent must be run on the solution of each subproblem, rather than just the overall solution.}  The main drawback of this algorithm, or any other which solves Eq.\eqref{eq:l0}, is they are NP-hard.  Their runtimes will always be exponential in $k$, making them intractable even for moderately sized problem.

The main benefit of this algorithm, and L0-norm algorithms in general, is that they can correctly solve for $x$ with $\hat{x}$ using only $k+1$ measurements\cite{foucart_mathematical_2013}.  $k+1$ is minimal, as there are $k$ degrees of freedom in the magnitudes of the non-zero elements and one extra to identify which elements are non-zero.  But this recovery is numerically unstable, as the number of bits of information needed to encode the positions of the non-zero elements is $O(k\log(n/k))$, the same needed for a run-length encoding.  By the pigeonhole principle, at least one of the measurements would need to encode $O(\log(n/k))$ bits, meaning we need arbitrarily precise and noiseless measurements.  Stable recovery requires a constant amount of information in each measurement, giving a lower bound of $O(k\log(n/k))$ measurements\cite{foucart_mathematical_2013}.

\subsection{L1 - Linear Programming}

L1-norm based algorithms such as Basis Pursuit\cite{shaobing_chen_basis_1994}, LASSO\cite{santosa_linear_1986}, and the Dantzig selector\cite{candes_dantzig_2007}, are a staple of compressive sensing literature.  The fundamental concept behind all of these algorithms is a relaxation of the NP-hard L0-norm of Eq.\eqref{eq:l0} into a linear constraint.  In this work, we solve the problem:
\begin{equation}
\label{eq:l1}
\begin{aligned}
\min_{\hat{x}} C\|y-A\hat{x}\|_{1} + \|w^\intercal\hat{x}\|_{1}
\quad 
\textrm{s.t.}
\quad 
\hat{x}_i \geq 0
\end{aligned}
\end{equation}
where $A$, $y$, and $\hat{x}$ are the same as in Eq.\eqref{eq:l0}, $C$ is a constant greater than 1 (we use $C=3$ in our experiments), and $w$ is a vector balancing the relative mass contributions of the elements of $x$, specifically $w_j = (\sum\limits_{i}A_{i,j})^{-1}$.  Eq.\eqref{eq:l1} is a convex relaxation of Eq.\eqref{eq:l0}\cite{foucart_mathematical_2013}.  Linearizing the problem through convex relaxation reduces its exponential runtime into something more tractable.  For example, Eq.\eqref{eq:l1} can be solved with a linear program in polynomial time.  The price of the relaxation is an increase in the number of measurements needed for recovery.  Although we lose the L0-norm's ability to recover using only $k+1$ high-precision measurements, the asymptotics for stable recovery stay exactly the same at $O(k\log(n/k))$ measurements.  In other words, L1-norm algorithms make compressive sensing tractable, while requiring only a constant factor times more measurements than the theoretical minimum.

Prior work extends L1 recovery to continuous or super-resolved bases, given some band-limit or other restriction on high-frequency variations\cite{fannjiang_super-resolution_2012,candes_towards_2014}.  Related work addresses the more general problem of coherent, or overcomplete, measurement matrices\cite{candes_compressed_2011}.  Candès et al. proved that multi-point source super-resolution is possible given that the sources are sufficiently separated\cite{candes_towards_2014}.  Their result motivates our use of a super-resolved measurement matrix, as well as our exclusion of test data containing sources within one spot size, the diffraction-limited resolution, of each other.  We find that our super-resolved, discrete recovery from Eq.\eqref{eq:l1} can be extended to produce continuous solutions by centroiding.  We calculate the centroids by center-of-mass.

\subsection{Data Collection Details}
The calibration (lookup table) for point source recovery is generated by scanning one of the two point sources independently (while the other is not illuminated) in a 2D calibration grid.  As the sources are indistinguishable, the calibration for one point source is valid for both, so only one source needs to be calibrated.  A second scan is performed with either one source or two sources simultaneously, to test whether the source positions can be recovered using the calibration scan.  In the two-source test, although only one point source is calibrated, the recovery for both point sources is done using the calibration.

For both the one-source and two-source tests, the calibration grid was collected with a single source immediately before the test grids to minimize the influence of environmental factors and drift.  The test and calibration grid spacings were chosen to have a least-common-multiple slightly larger than the field-of-view to evenly sample the interpolation regions between calibration points.  To prevent grating lobes from creating outliers, the calibration and test grids were restricted to 90\% of the field-of-view in each dimension: respectively, 60 and 42 samples per dimension for single source, and 84 and 16 samples per dimension for two source.  In total, this gave 1,764 test points for single source and 65,536 for two source.  After randomly trimming down the two source test points by a factor of 16 and removing points with sources within the spot size of each other, we were left with 3,851 two-source test points.  To prevent the camera from saturating, all data points used the same total amount of light in the sources.  In other words, for the two-source test points, each source had half as much light as the source for the single-source test points.

\begin{figure}[h]
	\begin{center}
		\includegraphics[height=6cm]{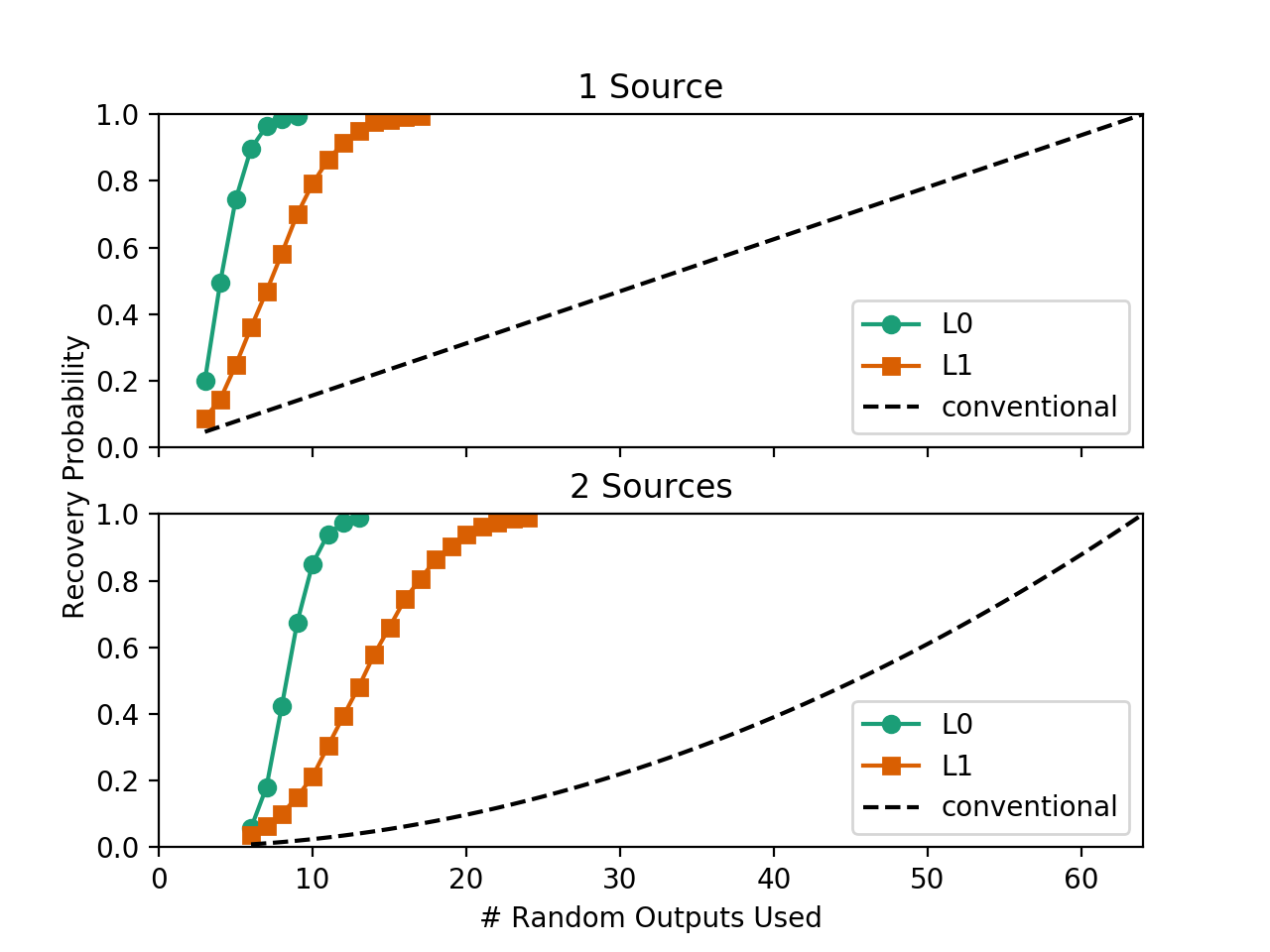}
	\end{center}
	\caption{\textbf{Algorithm Robustness} – Probabilities of recovering all sources to within half of the diffraction limited spot size (230 arcseconds) given a random subset of measurements. The performance of the canonical approximation algorithm, L1-relaxation, is compared with the optimal, but generally intractable, L0 algorithm. The theoretical performance of an equivalent conventional imager is shown for comparison: linear in pixels used for single source, and quadratic for two sources.}
	\label{fig:algorithm_robustness}
\end{figure}

\subsection{Experiment 1: Algorithm Robustness}

The first experiment evaluated how well the device functions as a compressive sensor.  Compressive sensors need fewer measurements, $y$, than conventional sensors to reconstruct the same sparse signal, $x$.  A conventional focusing imager needs at least as many measurements as there are antennas to guarantee recovery.  More generally, for a $k$-sparse signal and $n$ antennas, a focusing imager needs a number of measurements linear in $n$, $O(p^{1/k}n)$, to achieve a given recovery probability $p$.  In contrast, a compressive sensor can recover signals with a number of measurements only logarithmic in $n$, $O(k\log(n/k))$.  The robustness of a given compressive recovery algorithm determines the constant factor.  In other words, a more robust recovery algorithm needs fewer measurements to recover the same sparse signal.

For every test point, we randomly select a subset of the 64 device outputs as our measurements.  In Fig.~\ref{fig:algorithm_robustness}, we plot the ratio of test points with predicted source positions within 230 arcseconds, half of the spot size, of their actual positions.  For the L1 algorithm, when there are more centroids than sources, the brightest $k$ centroids are used.  To be able to centroid in two dimensions, a focusing imager needs at least 3 measurements per source, one per independent variable.  We use 3 and 6 measurements as minima for our one source and two source tests, respectively.  Additionally, only points with less than a 100\% recovery rate are shown to clearly distinguish them on the graph.  We see that the compressive sensing algorithms outperform the conventional baseline, with the L0 algorithm only needing 4 measurements to correctly recover 50\% of the single source test points.

To mitigate distortions caused by variations in the overlap of information in the measurements, we
compare the 90th percentiles between one and two source recovery.  L0 recovery reaches 90\%
recovery at 6.05 and 10.55 measurements for single and double source,
respectively, while L1 needs 11.72 and 18.90 measurements.  The ratios, 1.69 for L0 and 1.58 for
L1, are consistent with the theoretical 1.67 predicted by plugging in $k$ = 1, 2 and $n$ = 64 into $O(k\log(n/k))$.  The L1 algorithm needs almost twice as many measurements as L0 to achieve the same recovery rate.  This slightly larger constant factor is the performance tradeoff for L1's looser constraints and tractable runtime.

\begin{figure}[h]
	\begin{center}
		\includegraphics[height=7cm]{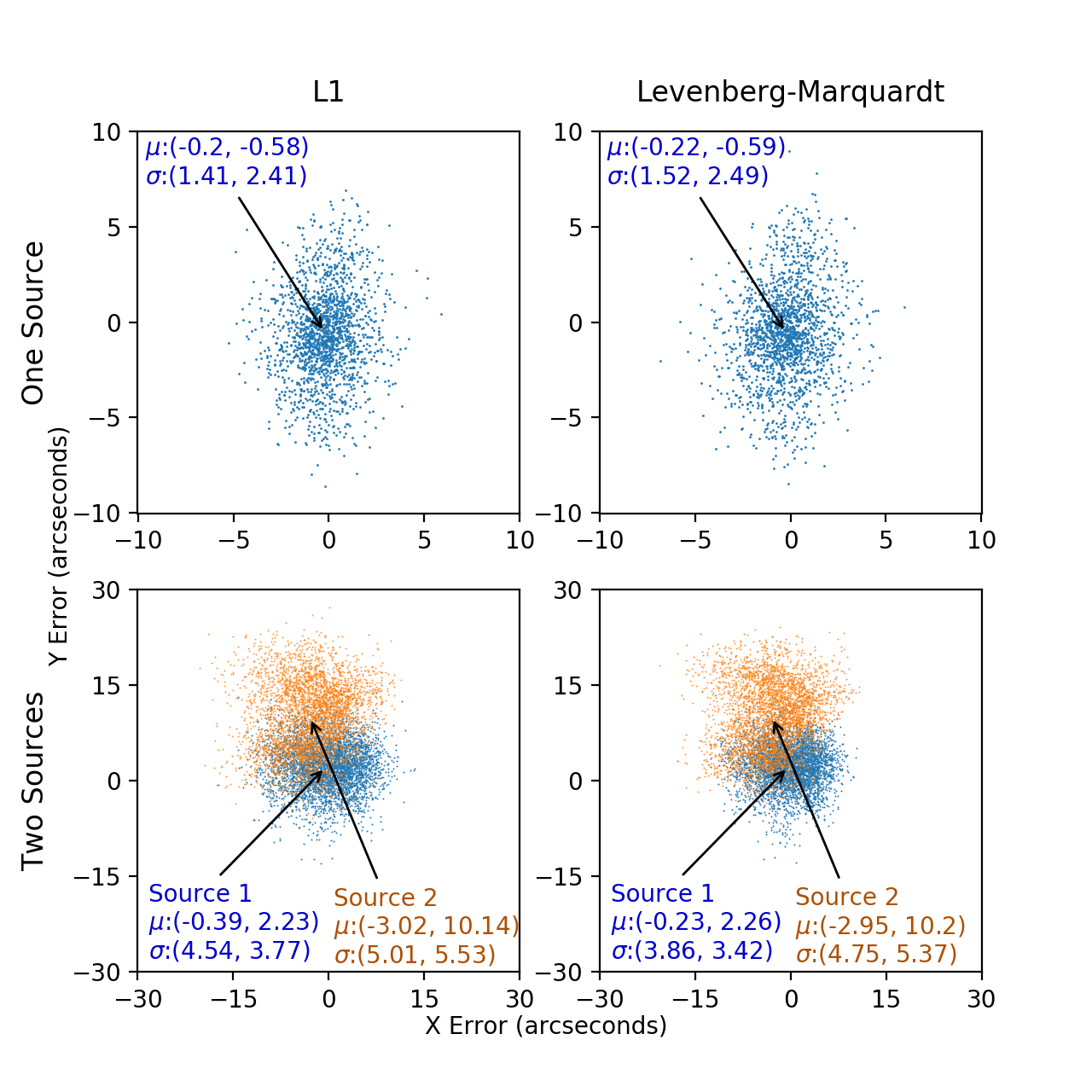}
	\end{center}
	\caption{\textbf{Recovery Precision} – L1’s centroiding precision is shown to be similar to a direct least-squared-error minimization, which is performed by hill-climbing interpolated point sources with the Levenberg-Marquardt algorithm, seeded by an oracle.  The $\mu$ and $\sigma$ shown are respectively the mean and standard deviation of each set of data.}
	\label{fig:recovery_precision}
\end{figure}

\subsection{Experiment 2: Multi-source Recovery Precision}

In the second experiment, we compared the centroiding accuracy of the L1 algorithm to the accuracy of a least-squared-error oracle receiver\cite{coluccia_exact_2014}.  Oracle receivers start with the knowledge of where the sources are located, so their performance is not impacted by misidentifications.  Oracle receivers represent the best an algorithm can hope to perform under the given assumptions.  In our experiment, the oracle receiver's minimization was performed by seeding the Levenberg-Marquardt algorithm with the actual locations of the sources.  The oracle receiver's search space was made continuous by interpolating the calibration grid with a Delaunay triangulation.

The results of the second experiment are shown in Fig.~\ref{fig:recovery_precision}.  The L1 algorithm performs similarly to the oracle receiver, with a precision of 2.8 arcseconds for the single source data and 7.5 arcseconds for two sources.  Both are well within the 5.0 \textmu m (13 arcsecond) guaranteed accuracy of the TRB25CC motors used in the experiment, suggesting the device may be able to achieve even higher levels of precision.  The 8.4 arcsecond offset between the first and second source's average recovered positions is an artifact of using the first source's calibration to recover both sources.  The sources were aligned by illuminating one fiber and positioning it to maximize the light reflected back into the second fiber by the beam splitter.  8.4 arcseconds corresponds to an offset of 3.1 \textmu m.

The device's grating lobe-free field-of-view is given by the small angle approximation-based equation: $\lambda/d$, where $\lambda$ is the center wavelength of 1545 nm and $d$ is the grating spacing of 100 \textmu m.  The field-of-view is then .0155 radians or 3,190 arcseconds.  With 2.8 arcseconds of precision for a single source over 90\% of the field-of-view, we have demonstrated a dynamic range of more than 1,000.

\section{Discussion}

Applying compressive sensing to silicon photonics opens the door for highly resilient imaging systems.  Using compressed sensing, calibration can take the place of phase trimming in accounting for phase errors.  In our device, we substitute software for complex integrated electronic components like thermal phase shifters, which require their own form of calibration to be able to account for fabrication-induced phase noise.\cite{poulton_optical_2016}  Coupled with the monolithic, flat nature of silicon photonics, we get a device that is not only resilient to manufacturing variations but also environmental challenges like vibrations, large forces, or temperature swings.  In contrast, it can be quite difficult to design such resilience into the supporting structures maintaining the focal region in conventional optics.

Over the life of an imager, components can become damaged or even lost entirely.  Especially in radiation-heavy environments, detectors often burn out.  When detector-loss happens in a lens-based imager, gaps can appear in its field-of-view.  But in a compressive sensing imager, source detection is distributed over a large number of outputs.  We showed that our device can still reliably resolve pairs of sources when only a quarter of its outputs remain functional.  Even when detectors are not known to be malfunctioning, which can create false positives in conventional systems, the linear weighting of outliers in our L1 algorithm still allows for stable recovery.

Silicon microphotonics is emerging as a promising platform for optical sensors.\cite{fatemi_88_2017,fatemi_nonuniform_2019,poulton_large-scale_2017,poulton_optical_2016}.  Separately, MMI-based compressive sensors are also being explored\cite{shin_single-pixel_2016,valley_multimode_2016,stork_imaging_2018,tragardh_label-free_2019}.  The novelty of our device is its ability to image passively, over a wide bandwidth, and without scanning.  Previous work in MMI-based, and in general speckle-based, compressive imagers has focused on using ghost imaging\cite{katz_compressive_2009}\cite{shin_single-pixel_2016}.  In ghost imaging, the MMI is used to project a speckle pattern onto the object, which reflects light back into a single-pixel detector.  Our device directly images using the MMI, allowing it to operate passively and without the need for multiplexing.

Imaging using silicon microphotonic phased arrays\cite{fatemi_88_2017,zhang_grating-lobe-suppressed_2018} has focused on narrowband devices, with single-pixel field of view with only a few broadband exceptions\cite{clevenson_incoherent_2020}.  Narrowband devices cannot be used for low, natural light imaging or high bandwidth communication.  To cover a large field-of-regard, single-pixel field-of-view devices need to raster across the object, typically through time-multiplexed scans.  In contrast, our device, without scanning, can localize a 40 nm bandwidth point source to one of over $10^{6}$ places in its field-of-view.

\section{Conclusion}

This work represents, to our knowledge, the first application of compressive imaging in a photonic-integrated device. The combination of broadband optical phased arrays and mixing components has allowed us to take a step towards significant volume reduction in optical compressive sensors.  Our device is fully passive, having no need for phase shifters, as measurement matrix calibration makes the measurements robust to phase errors. We demonstrated simultaneous source brightness recovery of two sources and localization with better than 10 arcsecond precision in a sub-millimeter thick form-factor.  For a single light source in the field of view, we achieve a recovery rate higher than 99.9\% using 10 of the 64 outputs, and a 90\% recovery rate with only 6 of the 64 outputs. This planar optical phased array compressive sensor is well-suited for imaging sparse scenes in applications constrained by form factor, volume, or high-cost detectors, with the potential to revolutionize endoscopy, beam locators, and LIDAR. Future work includes increasing the device area to increase resolution, optimizing to decrease component loss on the device, packaging the sensor for further testing and field environments, and moving into applications-based testing. 

% Appendix~\ref{sec:misc}

% Footnotes\footnote{Footnote example.} may be used

% A label of “Video/Audio 1, 2, …” should appear at the beginning of the caption to indicate to which multimedia file it is linked . Include this text at the end of the caption: \url{http://dx.doi.org/doi.number.goes.here}}

%\acknowledgments

% References
\bibliography{biblio3}
\bibliographystyle{spiebib} % makes bibtex use spiebib.bst

\end{document}